\documentstyle[sprocl]{article}
\begin{document}

\title{ON THE HEAVY  RELIC NEUTRINO - GALACTIC GAMMA HALO  CONNECTION }
\author{D. Fargion $^1$ ,
 R. Konoplich $^{2,3}$, M. Grossi $^1$, M. Khlopov $^{2,4}$}

\address{$^1$
Physics Dept. Rome University 1 and INFN,\\ Piazzale A. Moro 2,
00185 Rome, Italy.\\
 $^2$ Center for Cosmoparticle Physics - "Cosmion", Moscow,
Russia.\\ $^3$ Moscow Engineering Physics Institute.\\ $^4$
Institute of Applied Mathematics "M. V. Keldysh", Moscow, Russia.}

\maketitle

\begin{abstract}
A halo model with heavy relic neutrinos N belonging to a fourth
generation and their annihilations in galactic halo may explain
the recent evidence of diffused gamma (GeV) radiation around
galactic plane. We considered a neutrino mass in the narrow range
($M_Z/2 < m_N < M_Z$) and two main processes as source of gamma
rays. A first one is ICS of ultrarelativistic electron pair on IR
and optical galactic photons and a second due to prompt gammas by
$\pi^0$ decay, leading to a gamma flux ($10^{-7} \,-\, 10^{-6} \,
cm^{-2}\,s^{-1} \, sr^{-1}$) comparable to EGRET detection. Our
predictions are also compatible with the narrow window of neutrino
mass $45\, GeV < m_N < 60\, GeV $, required to explain the recent
underground DAMA positive signals.
\end{abstract}

\section{Introduction}

The recent observations of EGRET telescope show a diffuse $\gamma$
- ray emission in the halo of our galaxy [1]. Actually models
trying to explain a gamma halo range between\\ a) a high galactic
latitude distribution of high energy cosmic ray sources (fast
running pulsar "Geminga" like) [1][2],\\ b) collisions of high
energy protons (tens of GeV) in molecular clouds (mostly $H_2$),
whose formation is favored in galactic halos [3],\\ c) a cold dark
matter scenario with neutralino, the lightest supersymmetric
particle in the minimal SUSY model, whose annihilations in heavy
fermions ($c\bar{c},\,b\bar{b},\,t\bar{t}$) and bosons
($W^+W^-,\,ZZ,\,gg$) could lead to gamma secondaries emission
[4],\\ d) Inverse Compton Scattering (ICS) of interstellar photons
off cosmic ray electrons with a spectra harder than previous
predictions [5].\\ A heavy Dirac neutrino of a fourth generation
in a Cold Dark Matter (CDM) model, at masses $m_N > M_Z/2$, may
offer an elegant solution to the gamma ray signal detected, though
it can not solve all the dark matter problem in the halo.\\
Neutrino annihilations at high galactic latitude could produce\\
1) $\gamma$ rays by ICS of relativistic electrons (primaries or as
secondary decay products of heavier leptons in the annihilation
chains $N\bar{N} \rightarrow l^+l^- \rightarrow e^+ e^-$) onto
thermal photons ($IR$, $optical$) near and above the galactic
plane,\newline 2) direct gammas by neutral secondary pions
decay.\\ The estimated flux we derived here is roughly close to
EGRET results in the hypothesis of a smooth and homogeneous
galactic halo.\newline

\section{The heavy neutrino model}

LEP I  has fixed sever constrains on the number of "light"
($m_{\nu} \ll M_Z/2$) neutrino families from Z width data. However
there is no experimental prohibition on the existence of an
additional heavy neutral lepton with mass $M_N > M_Z/2$. Such a
heavy stable neutrino belonging to a fourth fermion family was
introduced nearly 20 years ago as a CDM candidate [6],[7]. An
apparently simple fourth generation model is not so easy to build.
Anyway models predicting a fourth heavy stable neutrino can be
found in [8],[9],[10]. In an expanding universe scenario heavy
neutrinos decouple from a thermal equilibrium condition when
global temperature drops below their rest mass energy ($T < m_N$)
and weak interactions become too slow to keep neutrinos in
equilibrium with the cosmological fluid. From this moment the only
change in neutrino density is due to cosmic expansion. The relic
abundance is given by [11]

\begin{equation}
\ n_N \simeq \frac{2 \times 10^{-18}}{g_*^{1/2} M_p m_N
(\overline{\sigma
\beta})_f} \bigg[ 40 + \ln \left( \frac{g_s}{g_*^{1/2}} M_p m_N (\overline{%
\sigma \beta})_f \right) \bigg] n_{\gamma} (T)
\end{equation}
\

where $g^* = N_{bos} + \frac{7}{8} N_{ferm}$ is the number of
effective degrees of freedom at temperature T, $g_s$ is the number
of particle spin states $M_p$ is the proton mass,
$(\overline{\sigma \beta})_f $ is the thermally averaged
annihilation cross section at freeze out, $n_{\gamma} = 0.24 T^3$
is the cosmic photon number density.\\ Annihilations of heavy
neutrinos in the universe happen through two main channels\\ {\bf
channel 1)} $ N\bar{N} \rightarrow f \bar{f}$ if $M_Z/2 < m_N <
m_W$ \\
where the cross section decreases as $4m^2_N/(4m^2_N - M^2_Z)^2$ for growing $m_N$ %
\newline
{\bf channel 2)} $N \bar{N} \rightarrow W^+ W^-$ if $m_N > m_W$\\
with a  cross section growing like $m^2_N$ for increasing $m_N$
[12].\newline As $\rho_N \propto \sigma^{-1}$, neutrino relic
density exhibits a maximum (with $\rho_{max}/\rho_c \approx
10^{-2} h^{-2}$) near $m_N \sim M_W$ and then starts to decrease
as $m_N^{-2}$ in the mass range $m_N > m_W$, without reaching the
critical value $\Omega = 1$, at least in the mass range where
Standard Model may be applied ($m_N < 1\, TeV$).\newline
Clustering during galactic structure formation determine an
increase in neutrino density [13],[14],[15] that in the central
part of the galaxy could be as large as 5 $\div$ 7 orders of
magnitude (the exact value depends on the other CDM, HDM densities
and masses) .\\ A spherical halo around our galaxy made of heavy
neutrinos is the model proposed in order to explain $\gamma$
emission observed at high galactic latitude. In a galactic halo,
neutrinos with a higher density distribution could annihilate
again leading to a flux of ordinary particles beyond the galactic
plane, potential sources of high energy radiation.\\ Constrains on
neutrino mass come either from cosmological data (not too high
pollution of $e^+ e^-$ cosmic rays is observed in the range $M_Z <
m_N < 300\,GeV$ [11]) or DAMA detector, where recent signals could
be attributed to heavy neutrino in the mass window $45\,GeV
< m_N < 50\, GeV$[15].

\section{Neutrino annihilation products as gamma ray source}

\subsection{Relativistic electron pairs: ICS on the galactic interstellar
radiation field(ISRF)}

Heavy neutrinos could directly annihilate in relativistic electron
pairs (either prompt ones or born through secondary decay
processes of heavier particles $\mu,\,\tau$, as in channel 1 way)
. Channel 2 leads to electron pairs through leptonic decay of W
($N\bar{N} \rightarrow W^{+}W^{-}\rightarrow l^+ l^- $). Electron
pairs may be generated even by W, Z hadronic decay through charged
pions and neutrons production.\\ Electrons and positrons are
trapped by galactic magnetic field, and propagating through the
Galaxy loose either $\,$''memory''$\,$ of their $\,$''place of
birth''$\,$ as well as energy for bremsstrahlung, synchrotron or
ICS. These processes determine a
broadening of different electron ''lines'' ($N\bar{N}\rightarrow l^+l^-%
\rightarrow e^{+}e^{-}$), so that $e^{-}\,(e^{+})$ spectra (even
considering electrons and positrons that come from hadron decays)
are at final stages described by the consequent approximated power
law $J=KE^{-\alpha }$ (where K is a normalization constant).
Numerical simulation of $N\bar{N}$ annihilation performed with the
package PYTHIA 5.7 [16] with suitable modifications to include a
fourth generation of fermions, show that such $N\bar{N}$ \ relic
electron fluxes are considerably lower than observed neighbor
galactic background (in the range of masses
$45\,GeV<m_{N}<M_{Z}$). Such a cosmic ray input can not be used to
confirm or refute heavy neutrino presence in galactic
halo.\newline ICS of "soft" background photons (isotropic CBR or
anisotropic infrared and optical interstellar radiation field) off
relativistic electrons created out of the galactic plane is a
possible source of radiation. Energies order of magnitude is fixed
by the ICS characteristic relation $E_{\gamma} = 4/3
\epsilon_{ph}(E_e/m_e c^2)^2$, where $\epsilon_{ph} $ is the
target photon energy.\newline We excluded here collisions with
microwave photons which would require too large neutrino masses
($m_N > 1\, TeV$). No clear theory is still available for such a
heavy particle. ICS on IR and optical photons needs respectively
$E_e \geq 50\,GeV$ and $E_e \geq 10\,GeV$, and is more efficient
in gamma ray production. The Galaxy is a disk-like radiative
source of radius $\sim$ 15 $kpc$, so the interstellar radiation
field has a vertical extent of several kpc. We assumed this
radiation to be represented by the obvious scaling law

\begin{equation}
\ n_{ph} (r) = \frac{n_{ph} (0)}{1 + r^2/a^2_{\gamma}}.
\end{equation}

where $r$ is the distance from the galactic plane, and $a_{\gamma}
= 10\,kpc$ is the characteristic length of interstellar radiation
distribution in the Galaxy. Photon density could be considered
roughly constant in a region of radius $a_{\gamma}$ [15].\newline
An electron distribution ($KE^{-\alpha}$) interacting by ICS with
photons at energy $\epsilon_{ph}$ and density $n_{ph}$ generate
radiation whose intensity is [17]

\begin{equation}
\ J_{\gamma} (E_{\gamma}) = \frac{2}{3} K a_{\gamma} n_{ph}
\sigma_T \left( \frac{\bar{\epsilon}_{ph}}{(mc^2)^2}
\right)^{(\alpha -1)/2} E_{\gamma}^{-(\alpha + 1)/2}
\end{equation}
\

where $n_{ph}$ is the target photon background density,
$\bar{\epsilon}_{ph}$ its average energy and $\sigma_T$ is the
Thomson cross section. \newline Gamma intensity has been
calculated for $m_N = 45,\, 50,\, 100,\,300\,GeV$. The largest
flux has been obtained for ICS on optical photons.\\ For $m_N = 50
\,GeV$ the calculated flux is

\begin{equation}
\ \frac{dN_{\gamma}}{dS\, dt\, d\Omega\, dE_{\gamma}} \simeq 2
\cdot 10^{-7}
A(\psi) \left( \frac{E_{\gamma}}{GeV} \right)^{-1.55} \left( \frac{a_{\gamma}%
}{10\, kpc} \right) \, cm^{-2}\,s^{-1}\, sr^{-1}\,GeV^{-1}
\end{equation}

and for $m_N = 100\,GeV$ one finds

\begin{equation}
\ \frac{dN_{\gamma}}{dS\, dt\, d\Omega\, dE_{\gamma}} \simeq 3
\cdot 10^{-7}
A(\psi) \left( \frac{E_{\gamma}}{GeV} \right)^{-1.5} \left( \frac{a_{\gamma}%
}{10\, kpc} \right) \, cm^{-2}\,s^{-1}\, sr^{-1}\, GeV^{-1} \,.
\end{equation}

$A( \psi )$ is the adimensional integral of interstellar photon
density along the line of sight L, defined by the angular
coordinate $\psi$ (angle between L and the direction of the
galactic centre). $A(\psi)$ is of few unities and corresponds to

\begin{equation}
\ A( \psi ) = \frac{1}{a_{\gamma}} \int_{line\, of\, sight} \frac{dr(\psi)}{%
(1 + r(\psi)^2/a^2_{\gamma})}
\end{equation}

Gamma intensity due to infrared background is less abundant than
optical photons as a consequence of the spectral power law
$E^{-\alpha}$.\newline Assuming an average $N\bar{N}$ clustering
$\rho_N^{gal}/ \rho_N^{cosm} = 10^6$, the flux obtained for two
values of $m_N$
\\ 1)$\Phi_{\gamma} (E > 1\, GeV) \simeq 4 \cdot 10^{-7}\,A(\psi)\, \left( \frac{%
a_{\gamma}}{10\, kpc} \right) \, cm^{-2}\,s^{-1}\,sr^{-1}$,$\;$
($m_N \simeq 50\,GeV$),\newline 2)$\Phi_{\gamma}
(E > 1\, GeV) \simeq 6 \cdot 10^{-7}\,A(\psi)\, \left( \frac{%
a_{\gamma}}{10\, kpc} \right) \, cm^{-2}\,s^{-1}\,sr^{-1}$ $\;$
($m_N \simeq 100\,GeV$),\\ is comparable with EGRET
observations:\\ $\Phi_{\gamma} (E > 1\, GeV) \simeq 8 \cdot
10^{-7} \, cm^{-2}\,s^{-1}\, sr^{-1}$.\\ Additional tests of this
model could be obtained with the nearly detectable signal of
hundreds KeV radiation in the halo with flux $J_{\gamma} \simeq
10^{-2} \, cm^{-2}\,s^{-1}\, sr^{-1}$ at peak energy $E_{\gamma}
\sim 300 keV$ as well as a flux $J_X \simeq 0.3 \,
cm^{-2}\,s^{-1}\, sr^{-1}$ at $E_X \sim 3 \,KeV$ due to ICS of
tens of $GeV$ and $GeV$ electrons with CBR. An addtional parassite
radio background arises at high galactic latitudes due to
synchrotron losses of the same electrons at $E_e \sim 10\,GeV$,
with typical density flux

\begin{equation}
\ J_{sync} \sim 5 \cdot 10^4 \left( \frac{B}{1 \mu G} \right ) \left( \frac{%
\gamma}{2 \cdot 10^4} \right)^{-2} \left( \frac{U_{rad}}{0.2 \,
eV\, cm^{-3}} \right)^{-1} \,Jy \,.
\end{equation}
\

and average frequency

\begin{equation}
\ \nu = \gamma^2 \left( \frac{eB}{2 \pi m_e} \right) \sim 1\, GHz \left(
\frac{B}{1 \mu G} \right) \left( \frac{\gamma}{2 \cdot 10^4} \right)^2
\end{equation}
\

where we used as characteristic scale for the magnetic field B = 1
$\mu G$,and as optical background energy density $U_{rad} = 0.2 \,eV\,cm^{-3}$.%
\newline

\subsection{Annihilations in gamma photons}

Gamma radiation could also be produced in neutrino annihilations
due to neutral pions secondaries in Z, W hadronic decay. This kind
of emission does not need to introduce any kind of radiative
background and is directly related to neutrino distribution in the
halo.\\ Photon flux is described by the following expression:

\begin{equation}
\ J_{\gamma} = \frac{1}{4 \pi m^2_N} \sum_{i} \sigma_i v \frac{dN^i} {dE}
\int_{line\,of\,sight} \rho^2 \,(r) dr(\psi)
\end{equation}
\

where $\psi$ is the angle between the line of sight and the galactic center,
$\rho\,(r)$ is heavy neutrino density as a function of galactocentric
radius, and $\sum_i \sigma_i v \frac{dN^i}{dE}$ counts all possible final
photon channels ($\frac{dN^i}{dE}$) which could contribute to gamma photons
emission. The integral of neutrino density along the line of sight L depends
on the halo model chosen for dark matter distribution, which is generally
described as

\begin{equation}
\ \rho\,(r) \propto \frac{1}{(\frac{r}{a})^{\gamma} [1 + (\frac{r}{a}%
)^{\alpha} ] ^{(\beta - \gamma)/\alpha}}
\end{equation}
\

The simplest density profile is described by an isothermal sphere
with $\alpha= 2 ,\,\beta= 0,\, \gamma=0$) and length scale $a \geq
10\,kpc$.\\ In the spherical model the square density integral
leads to an adimensional intensity $I(\psi)$

\[
I(\psi) = \int \frac{1}{(1 + (r/a)^2)^2} dr(\psi)/a ;
\]

this intensity has a characteristic behaviour which is maximum in
the direction of the galactic center ($\psi = 0$), and then
decreases for $0 < \psi < \pi$, but it doesn't vary more than a
factor ten with the angular coordinate.\newline At high latitudes
$I\,(\psi)$ is generally of order unity.\newline Models with a
singular behaviour towards the galactic center or which postulate
a clumpy distribution of dark matter could contribute to enhance
the total gamma flux in the halo, but we shall neglect them
here.\\ Monte Carlo simulations of neutrino annihilations [16]
have been also used to compare EGRET flux, showing that it is
possible to extrapolate a power law for gamma spectrum.\newline An
approximated integral flux for $m_N = 50\,GeV$ and $a \sim
10\,kpc$  is roughly

\begin{equation}
\ \Phi_{\gamma} > 6\cdot 10^{-7}\,I(\psi) \,cm^{-2} \,s^{-1} \,sr^{-1},
\end{equation}
\

while for $m_N = 100\,GeV$ at $a \sim 10\,kpc$

\begin{equation}
\ \Phi_{\gamma} > 4\cdot 10^{-7}\,I(\psi) \,cm^{-2} \,s^{-1} \,sr^{-1},
\end{equation}
\

\bigskip
In conclusion there are at least two independent processes able to
solve the puzzle of a GeV gamma halo by the role of a cosmic relic
DM made of fourth generation heavy neutrinos. The neutrino masse
are constrained into a narrow energy window ($45\,GeV < m_N <
60\,GeV$), in order to combine at once the DAMA data and the other
underground detector. This reality will be soon confirmed or
excluded by LEP II search for $e^-e^+ \rightarrow N\bar{N} \gamma$
events [19] in this energy band.


\end{document}